\documentclass[a4paper,11pt]{article}
\usepackage{jheppub} 
\usepackage{lineno}

\usepackage{amsmath,amssymb,mathtools} 
\usepackage{color}
\usepackage{graphicx}
\usepackage{subfigure}
\usepackage{caption}        
\usepackage{hyperref}            
\usepackage{multirow,makecell}   
\usepackage{textcomp}
\usepackage{wasysym}
\usepackage{setspace}
\usepackage{verbatim}
\usepackage{float}
\usepackage{ulem}
\usepackage{mhchem}
\usepackage{bm}
\usepackage[utf8]{inputenc}

\title{(In)stability of symbiotic vortex-bright soliton in holographic  immiscible binary superfluids }

\author[a,b]{Yu-Ping An}%
\author[a,b,c]{Li Li}
 \affiliation[a]{CAS Key Laboratory of Theoretical Physics, Institute of Theoretical Physics,
Chinese Academy of Sciences, Beijing 100190, China}
\affiliation[b]{School of Physical Sciences, University of Chinese Academy of Sciences, Beijing 100049, China}
\affiliation[c]{School of Fundamental Physics and Mathematical Sciences, Hangzhou Institute for Advanced Study, University of Chinese Academy of Sciences, Hangzhou 310024, China}
\emailAdd{anyuping@itp.ac.cn, liliphy@itp.ac.cn}

\abstract{ Symbiotic vortex-bright soliton structures with non-trivial topological charge in one component are found to be robust in immiscibel two-component superfluids, due to the effective potential created by a stable vortex in the other component. We explore the properties of symbiotic vortex-bright soliton in strongly coupled binary superfluids by holography, which naturally incorporates finite temperature effect and dissipation. We show the dependence of the configuration on various parameters, including the winding number, temperature and inter-component coupling. We then study the (in)stability of symbiotic vortex-bright soliton by both the linear approach via quasi-normal modes and the full non-linear numerical simulation. Rich dynamics are found for the splitting patterns and dynamical transitions. Moreover, for giant symbiotic vortex-bright soliton structures with large winding numbers, the vortex splitting
instability might be rooted in the Kelvin-Helmholtz instability. We also show that the second component in the vortex core could act as a stabilizer so as to suppress or even prevent vortex splitting instability. Such stabilization mechanism opens possibility for vortices with smaller winding number to merge into vortices with larger winding number, which is confirmed for the first time in our simulation.}

\begin{document}
\maketitle
\flushbottom

\section{Introduction}
Superfluidity, characterized by the flow of matter without resistance, represents a remarkable macroscopic quantum phenomenon observed in various systems, including ultracold atomic gases, liquid helium and nuclear matter in neutron stars. At the heart of superfluid dynamics lies vortices that are topological defects emerging from the quantization of circulation. These quantized vortices are characterized by a singular core around which the phase of the superfluid order parameter winds by $2\pi S$ with the winding number $S$ an integer, giving rise to nontrivial topological properties. Theoretical descriptions of vortex dynamics in superfluid systems have advanced significantly, driven by a combination of analytical insights and numerical simulations. The mean-field Gross-Pitaevskii equation (GPE) provides a useful framework for modeling the macroscopic behavior of weakly coupled superfluids in the absence of a viscous normal fluid component, and predicts the formation and evolution of vortices. Nevertheless, it is difficult to incorporate finite temperature effect and dissipation. Meanwhile, holographic superfluids offer us a powerful way to explore superfluidity in the strong coupling limit, together with finite temperature and dissipation effects~\cite{PhysRevLett.101.031601,Herzog:2008he}.

In single component superfluids, multiply quantized vortices with the winding number $|S|>1$ generally present splitting instability. Dynamics and splitting instability of vortices in single component case have been widely studied, both in weakly coupled Bose-Einstein condensates (BEC)~\cite{RN400,RN399,RN352,RN401} and strongly coupled holographic
superfluids~\cite{RN359,RN357,RN356,RN393,RN391,RN358}. The study of vortices in two-component superfluids has emerged as a frontier of research, presenting a fertile ground for uncovering novel physical phenomena and elucidating the interplay between different quantum phases~\cite{RN349,RN395,RN361,RN397,RN350,RN351,RN365,RN394,RN348,RN412,RN411,RN398}. In particular, including a second component could enrich the splitting instability of multiply quantized vortices. For immiscible binary superfluids, the second component can reside in the core of the vortex in the first component, acting as a stabilizer that suppresses or even eliminates the splitting instability. Such structure is called symbiotic vortex-bright soliton~\cite{RN350}, since the bright soliton in the vortex core can not exist alone for repulsive self-interaction. When the radius of the vortex-bright soliton is much larger than the interface width between the two components, such system can also be regarded as a two-fluid system with relative velocity, since the first component is rotating around the second component in the vortex core. Therefore, the Kelvin-Helmholtz instability (KHI) is also expected~\cite{RN312}.

Nevertheless, these studies of vortices in two-component superfluids mostly rely on GPE, which is only applicable in the weak coupling limit. Moreover, no finite temperature effects and dissipation have been considered. As shown in the recent work~\cite{2comKHI,An:2024dkn}, those effects could play an important role in the interface dynamics of strongly interacting immiscible binary superfluids. Distinct features compared to the traditional GPE have been uncovered.
In particular, the interface instability for two superfluid components moving with identical velocity was confirmed, thanks to the normal fluid components at finite temperature~\cite{An:2024dkn}. This feature is in sharp contrast to the prediction from GPE for which no such co-flow instability develops in an isolated uniform system because of Galilean invariance. In this paper, we use holographic two-component superfluid model to study the properties of symbiotic vortex-bright soliton structure.
We uncover dynamical instability of the system using both the linear perturbation analysis and full numerical simulation. We show that the window of instability are generically alleviated/suppressed by the increasing presence of the second bright component. Both the splitting and merging patterns are observed from the full non-linear numerical simulations. In particular, the later process, to the best of our knowledge, has not been reported in the literature.

This paper is organized as follows. We present our holographic model and setup in Section~\ref{sec2} and construct the static vortex-bright soliton in Section~\ref{sec3}. In Section~\ref{sec4} we use both linear analysis and full time evolution to study the (in)stability of symbiotic vortex-bright soliton structure for different temperatures and coupling strength. In Section~\ref{sec5} we consider the case when the radius of the composite vortex is much larger than the interface width between the two components and show the possibility for KHI. We conclude in Section~\ref{sec6} with some discussions. Technical details are included in Appendix~\ref{A} for time evolution scheme and in Appendix~\ref{B} for numerical computation of quasi-normal modes.

\section{Holographic model and setup}
\label{sec2}
In this work we consider a (3+1) dimensional holographic model for binary superfluids at strong coupling. At finite temperature, each of the superfluids contains both superfluid and normal viscous fluid components. For simplicity, we consider the case where the fluctuations of the temperature and the normal fluid’s velocity are frozen. This corresponds to the probe limit in the bulk description by neglecting the back-reaction of matter content to the geometry. To be specific, the Lagrangian for matter reads
    \begin{equation}\label{action}
        \begin{aligned}
             \mathcal{L}=&-(\mathcal{D}_\mu\Psi_1)^*
            \mathcal{D}^\mu\Psi_1-m_1^2|\Psi_1|^2-(\mathcal{D}_\mu\Psi_2)^*
            \mathcal{D}^\mu\Psi_2-m_2^2|\Psi_2|^2-\frac{\nu}{2}|\Psi_1|^2|\Psi_2|^2-\frac{1}{4}F^{\mu\nu}F_{\mu\nu},
          \end{aligned}  
    \end{equation}
where $\mathcal{D}_\mu\Psi_i=(\nabla_\mu-ie_iA_\mu)\Psi_i$, $A_\mu$ is the $U(1)$ gauge field, and $F_{\mu\nu}$ is the field strength. Two complex scalars $\Psi_1$ and $\Psi_2$ are charged under the $U(1)$ gauge field, corresponding to the two superfluid components of the dual system. The inter-component coupling $\nu$ determines the miscibility of the superfluids. We consider immiscible binary superfluids by taking $\nu>0$ to consider the interface dynamics~\cite{2comKHI,An:2024dkn}. Equations of motion can be found in Appendix~\ref{A}. 

As for background, we use a planar AdS black hole in Eddington-Finkelstein coordinates:
    \begin{equation}
        ds^2=\frac{L^2}{z^2}(-f(r)dt^2-2dtdz+dr^2+r^2d\theta^2),\quad f(r)=1-\frac{z^3}{z_h^3}\,,
    \end{equation}
where $z=0$ represent the AdS boundary and $z=z_h$ denotes the event horizon of the black hole. The background corresponds to a heat bath with temperature $T=3/(4\pi z_h)$ of the boundary system. To study the symbiotic vortex-bright soliton, we have considered polar coordinates $(r,\theta)$ instead of rectangular coordinates $(x,y)$ on the dual (2+1) dimensional boundary.

Without loss of generality, we set $L=z_h=1$ and choose the radial gauge $A_z=0$. For convenience, we mainly consider the case of the identical superfluids with $m_1^2=m_2^2=-2$ and $e_1=e_2=1$.\footnote{We shall also discuss the non-identical superfluids with $e_1\neq e_2$ in Section~\ref{sec5}.} The asymptotic expansions for $A_\mu$ and $\Psi_i$ near the AdS boundary $z=0$ read 
    \begin{equation}
        \begin{aligned}
            A_\mu&=a_\mu+b_\mu z+\mathcal{O}(z^2)\,,\\
            \Psi_i&=\Psi^{(v)}_i z^2+\mathcal{O}(z^3),\quad i=1,2\,.
        \end{aligned}
    \end{equation}
where we have taken the leading source terms for $\Psi_1$ and $\Psi_2$ to be vanishing to break the $U(1)$ symmetry spontaneously, \emph{i.e.} the superfluid state. Then $\Psi^{(v)}_i$ corresponds to the superfluid condensate $\mathcal{O}_i $. Moreover, the coefficients $a_t$ and $\bm{a}=(a_r, a_\theta)$ are interpreted as the chemical potential $\mu$ and vector potential, respectively. The latter is related to the superfluid velocity $\bm{v}=\nabla\theta_i-\bm{a}$ where $\theta_i$ is the phase of the superfluid condensate $\mathcal{O}_i$. In practice, we choose $\bm{a}=0$ and thus the superfluid velocity reads $\bm{v}=\nabla\theta_i$ for each superfluid sector. Furthermore, the charge conservation equation on the boundary is given by $-\partial_t b_t=b_i+(\nabla_i a_t-\partial_t a_i)$, with $-b_t=\rho$ the charge density. 

Thanks to the scaling symmetry of the system, $T$ and $\mu$ are not independent quantities. After fixing $z_h=1$, $\mu$ is the only free parameter. When $\mu\ge \mu_c\simeq 4.064$,\footnote{This critical value is obtained for the homogeneous and isotropic superfluid state with single condensate $\mathcal{O}_1$, \emph{i.e.} $\Psi_2=0$. After turning on the superfluid velocity that breaks isotropy, the critical chemical potential will increase and the superfluid phase transition will becomes a first order one when the superfluid velocity is beyond a threshold value~\cite{Herzog:2008he}.} the superfluid condensation develops a nonzero expectation value spontaneously via a second order phase transition, driving the system into a superfluid phase. This also fixes the ratio $T/T_c=\mu_c/\mu$ with $T_c$ the critical temperature below which the superfluid phase develops spontaneously.

\section{Static vortex-bright soliton configuration}
\label{sec3}
We begin with the static vortex-bright soliton. Thanks to the rotation symmetry, the bulk configuration is independent of the angular coordinate $\theta$ (except for the phase with a non-trivial winding number). The bulk ansatz reads
\begin{equation}\label{vbsoliton}
     \begin{aligned}          &\Psi_i=z\psi_i(z,r)e^{i\Theta_i(z,r,\theta)}, \quad \Theta_i(z,r,\theta)= S_i\theta+\Theta(z,r)\,,
          \\ & A_t=A_t(z,r), \quad A_\theta=A_\theta(z,r)\,, 
     \end{aligned} 
\end{equation}
where $S_i$ is an integer winding number of quantized vortex in the $i$-th component. The phase $\theta_i$ of the superfluids condensate $\mathcal{O}_i$ can be read off from $\theta_i=\Theta_i(z=0)$. In the present work we focus on the static vortex-bright soliton with $S_1=S$ and $S_2=0$, \emph{i.e.} no vortex exists in the second component. Then the equations of motion can be written as
\begin{equation}
\begin{aligned}   \partial_z(f\partial_z\psi_i)+\partial_r^2\psi+\frac{1}{r}\partial_r\psi+(\frac{A_t^2}{f}-\frac{(A_\theta-S_i)^2}{r^2}-z-\frac{\nu}{2}\psi_j^2)\psi_i=0\,, \\(i,j=1,2,\quad i\ne j)\,,
\end{aligned}
\end{equation}
\begin{equation}   f\partial_z^2A_t+\partial_r^2A_t+\frac{1}{r}\partial_rA_t-2A_t\sum_i\psi_i^2=0\,,
\end{equation}
\begin{equation}    \partial_z(f\partial_zA_\theta)+\partial_rA_\theta-\frac{1}{r}\partial_rA_\theta-2\sum_i(A_\theta-S_i)\psi_i^2=0\,.
\end{equation}
Note that the equations of motion for $\Theta$ and $A_r$ have be eliminated by setting $\partial_z\Theta=-A_t/f$ and $A_r=\partial_r\Theta$. 

Boundary conditions are given as follows. At the AdS boundary, we have
\begin{equation}
\begin{aligned}
    \psi_i|_{z=0}=0, \quad A_t|_{z=0}=\mu, \quad A_\theta|_{z=0}=0\,.
\end{aligned}    
\end{equation}
On the event horizon $z = z_h$, the regular boundary conditions are imposed, in particular, $A_t|_{z=z_h}=0$. Along the radius of the boundary direction, the system is cut off at a large radius $r=R_0$ where the Neumann boundary conditions are adopted:
\begin{equation}
    \partial_r\psi_i|_{r=R_0}=0,\quad \partial_rA_t|_{r=R_0}=0,\quad\partial_rA_\theta|_{r=R_0}=0\,.
\end{equation}
The cutoff $R_0$ is chosen to be sufficiently large compared to the vortex size such that the intrinsic dynamics of our vortex is not affected. At the vortex center $r = 0$, we impose the following boundary condition.
\begin{equation}
    \psi_1|_{r=0}=0,\quad\partial_r\psi_2|_{r=0}=0,\quad\partial_rA_t|_{r=0}=0,\quad\partial_rA_\theta|_{r=0}=0\,.
\end{equation}
Note that we have taken the Neumann boundary condition for $\psi_2$ at $r=0$ to allows a finite condensate $\mathcal{O}_2$ at the vortex center. Then the vortex-bright soliton configuration can be obtained numerically by using \emph{e.g.} Newton-Raphson method.

\begin{figure}[htpb]
        \centering
            \includegraphics[width=1\linewidth]{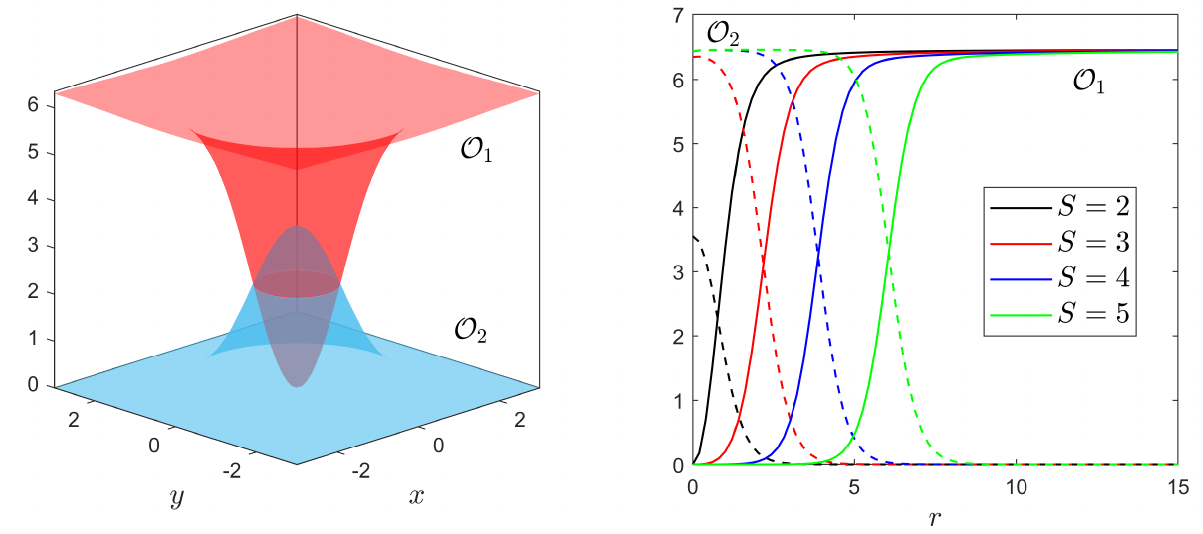}
            \caption{Illustration of the configuration for vortex-bright soliton in the holographic immiscible binary superfluids. \textbf{Left panel:} Vortex-bright soliton profile for the winding number $S=2$. Both the condensates $\mathcal{O}_1$ (red) and $\mathcal{O}_2$ (blue) are completely symmetric about the vertical at the core $r=0$ ($x=y=0$ in Cartesian coordinates). At the core, the value of $\mathcal{O}_1$ is vanishing while the one of $\mathcal{O}_2$ is maximum, which is common for other winding numbers. \textbf{Right panel:} Radial profiles of $\mathcal{O}_1$ (solid lines) and $\mathcal{O}_2$ (dashed lines) for different winding numbers. Radius of the vortex grows fast as the winding number $S$ increases. Relevant parameters are $T/T_c=0.677$ (or equivalently, $\mu=6$) and $\nu=5$. }
    \label{O-S}
\end{figure}

As a demonstration, in the left panel of Figure~\ref{O-S} we show the profiles of the superfluid condensates $\mathcal{O}_1$ (red) and $\mathcal{O}_2$ (blue) for the winding number $S=2$. It is clear that the first condensate $\mathcal{O}_1$ is vanishing at the core and increases along the radial direction, while the second one $\mathcal{O}_2$ takes maximum value at the core and decreases monotonically. This feature is common for vortex-bright solitons with different winding numbers.
The radial configuration is shown in the right panel of Figure~\ref{O-S} for different winding numbers. One can see that the radius of the composite vortex grows fast as $S$ increases, see in particular the solid curves of $\mathcal{O}_1$. This is a natural results since the centrifugal force grows as $S$, and therefore the circulation velocity $v=S/r$ increases. Meanwhile, the second condensate $\mathcal{O}_2$ (dashed curves) occupies larger region around the core. In particular, the peak value of $\mathcal{O}_2$ at the core increases and saturates quickly as $S$ is increased.

\begin{figure}[htpb]
        \centering
            \includegraphics[width=0.65\linewidth]{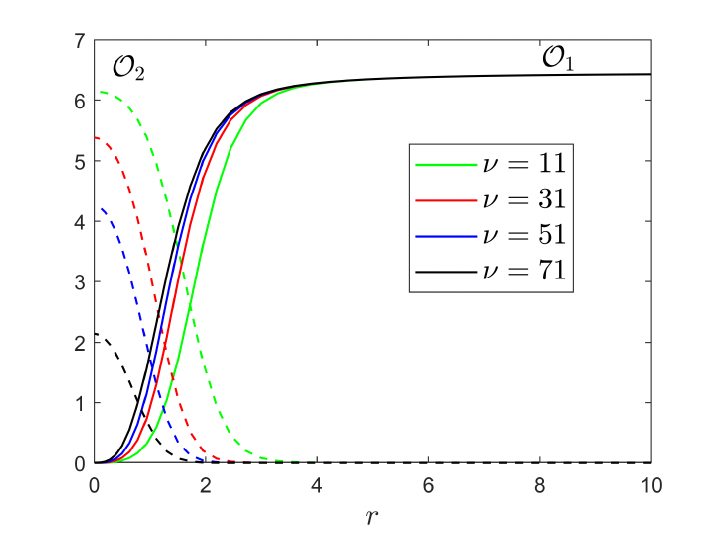}
            \caption{Profiles of the triply quantized composite vortices ($S=3$) for different $\nu$ at $T/T_c=0.677$. Both the radius of the vortex and the maximum value of $\mathcal{O}_2$ decrease as $\nu$ increases.}
    \label{O-nu}
\end{figure}

In Figure~\ref{O-nu}, we show the order parameters of the composite vortices with $S=3$ for different $\nu$ at $T/T_c=0.677$. We can see that as $\nu$ increases, the size of the triply quantized composite vortex shrinks and the maximum value of $\mathcal{O}_2$ at the core decrease. Thus, the inter-component coupling suppresses the formation of vortex-bright soliton. In contrast, the value of $\mathcal{O}_1$ far away from the core is independent of the coupling $\nu$. This feature can be understood as follows. The condensate $\mathcal{O}_2$ is exponentially suppressed far away from the core. Thus there exist only $\mathcal{O}_1$ for sufficiently large $r$, for which the system is effectively described by the holographic model with $\Psi_1$ only and the coupling $\nu$ plays no role. As is shown in Figure~\ref{O2-nu}, the value of condensate $\mathcal{O}_2$ at the core (solid black curve) decreases monotonically and eventually goes to zero beyond a critical coupling $\nu_c\approx77$. Meanwhile, the profile converges to that of triply quantized vortex with empty core as $\nu$ increases.

\section{(In)stability of symbiotic vortex-bright soliton}\label{sec4}
We now turn to the (in)stability of symbiotic vortex-bright soliton structure. In this section, we shall consider the triply quantized composite vortex which could exhibit the splitting instability of $p$-fold rotational symmetry with $p=2,3,4$.\footnote{The vortex of the winding number $S>2$ generically exhibits the splitting instability of $p$-fold rotational symmetry with $p=2,3,\cdots, 2(S-1)$.} Due to the narrow region of the scattering length as well as the tiny value of the growing rate, splitting patterns with $p\ge 4$ have not been observed experimentally. Moreover, multiply quantized vortices with $S>2$ share similar behaviors. Therefore, the triply quantized composite vortex serves as the most economical representative for illustrating the instability and the splitting patterns of multiply quantized vortices. We will first uncover the dynamical instability using the linear perturbation analysis. 
The linear approach will be then corroborated by numerical solutions to the full equations of motion.

\subsection{Linear instability via quasi-normal modes}
Once the stationary solutions are given, one can use the linear perturbation analysis via the quasi-normal modes. More precisely, one turns on perturbations on the background of the stationary configuration in Section~\ref{sec3}.
\begin{equation}
\Phi_i=\Phi_{i0}+\delta\Phi_i,\quad A_t=A_{t0}+\delta A_t,\quad A_r=A_{r0}+\delta A_r,\quad A_\theta=A_{\theta 0}+\delta A_\theta\,,
\end{equation}
with subscript $0$ denoting the background shown \emph{e.g.} in Figure~\ref{O-S}. We have introduced $\Phi_i=\Psi_i/z$ for numerical convenience. Thanks to the time translation symmetry and rotation symmetry of the vortex-bright solitons, one can take the perturbations with the following form.
    \begin{equation}\label{pfold}
    \begin{aligned}
        &\delta\Phi_i=u_i(z,r)e^{-i(\omega t-p\theta)}e^{iS_i\theta}, \quad \delta\Phi_i^*=v_i(z,r)e^{-i(\omega t-p\theta)}e^{-iS_i\theta},\\&\delta A_t=a_t(z,r)e^{-i(\omega t-p\theta)},\quad\delta A_r=a_r(z,r)e^{-i(\omega t-p\theta)}, \quad\delta A_\theta=a_\theta(z,r)e^{-i(\omega t-p\theta)}\,,
    \end{aligned}
    \end{equation}
with $p$ a constant and $S_i$ the winding number for the background. The excitation mode of~\eqref{pfold} has $p$-fold symmetry. Explicit form of the linearized  equations of motion can be found in Appendix~\ref{B}. For each $p$-mode, we can solve the generalized eigenvalue problem numerically and get the quasi-normal mode frequency $\omega$. Whenever the imaginary part of $\omega$ is positive, \emph{i.e.} Im$\omega>0$, this specific $p$-channel is dynamically unstable, triggering the instability of the symbiotic vortex-bright soliton.

\begin{figure}[htpb]
        \centering
            \includegraphics[width=0.65\linewidth]{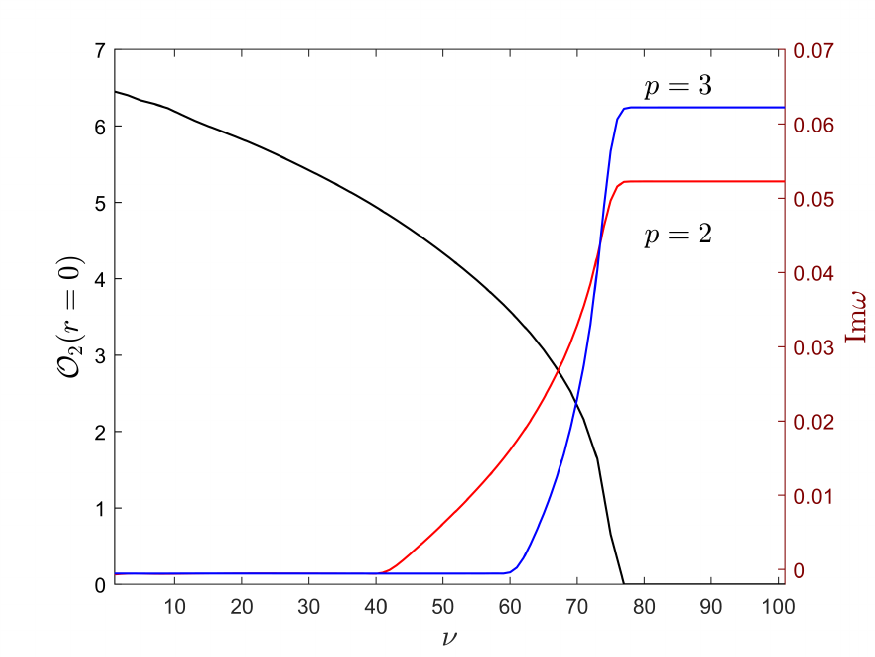}
            \caption{Maximum value of the second order parameter $\mathcal{O}_2(r=0))$ and $\mathrm{Im}\omega$ of $p=2$, 3 versus $\nu$ for $S=3$ at $T/T_c=0.677$. As $\nu$ increases, $\mathcal{O}_2$ (black curve) consistently decreases to zero, and the dominating unstable mode transits from $p=2$ to $p=3$ at $\nu\approx 73$. 
            }
    \label{O2-nu}
\end{figure}
Splitting patterns of $S=3$ vortex in holographic superfluid with single component (\emph{i.e.} by setting $\Psi_2=0$ in our present model) has been recently studied in~\cite{RN393}, where the authors found that the vortex is generically unstable at all temperatures for $p=2$, $3$ channels and unstable for $p=4$ channel only at relative high temperatures. Moreover, there is a transition of the dominating unstable channel from being $p=2$ to $p=3$ as the temperature increases. In Figure~\ref{O2-nu}, we present the imaginary part of the excitation frequency Im$\omega$ of $p=2$ (red curve) and $p=3$ (blue curve) channels for different $\nu$ at $T/T_c=0.677$. When $\nu\lesssim 40$, Im$\omega$ is consistently zero, which means such composite vortex is stable for these parameters. As $\nu$ increases, Im$\omega$ becomes positive, implying the onset of dynamical instability. According to~\cite{RN393}, for a $S=3$ vortex in holographic superfluid with $\Psi_2=0$, the mode with $p=3$ dominates the instability when $T/T_c=0.677$. However, turning on the second superfluid component $\Psi_2$ significantly changes this result. As visible from Figure~\ref{O2-nu}, as $\nu$ increase, it is the $p=2$ channel that becomes unstable first and dominates the system. The $p=3$ channel develops and eventually dominates the system only when the core value of $\mathcal{O}_2$ is close to zero, which means the core of the vortex is almost empty.

\begin{figure}[htpb]
        \centering
            \includegraphics[width=1\linewidth]{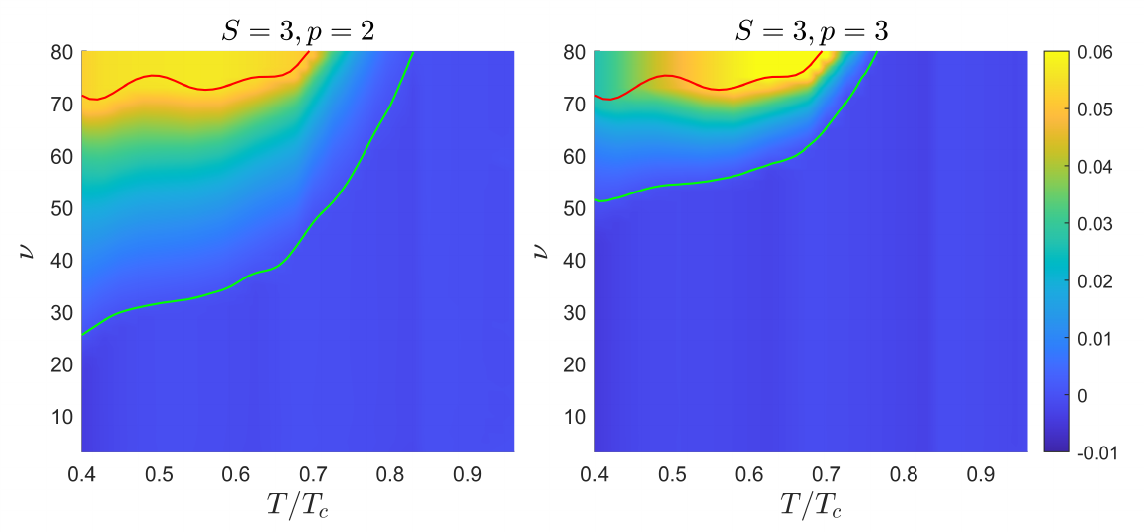}
            \caption{Spectrum of the low lying quasi-normal modes in terms of inter-component coupling and temperature for excitation modes with 2-fold symmetry (left panel) and 3-fold symmetry (right panel) of $S=3$ vortex-bright soliton. The green lines mark the boundary between the stable and unstable regions. Above the red line, the second superfluid component in vortex core is completely eliminated, \emph{i.e.} $\mathcal{O}_2=0$, for which the configuration becomes essentially a single component vortex.}
    \label{S=3}
\end{figure}
Figure~\ref{S=3} illustrates the density plot of Im$\omega$ in terms of $\nu$ and $T/T_c$ for $p=2$ channel (left panel) and $p=3$ channel (right panel).
The green lines mark the boundary between the stable (Im$\omega<0$) and unstable (Im$\omega>0$) regions. When $\nu$ is small, no instability exists at any temperatures due to the stabilizing mechanism of the second component in the vortex core. The coupling strength needed for the onset of instability increases as the temperature increases. This can be attributed to the growth of radius of vortex together with temperature (see the blue curve in the right panel of Figure~\ref{R-v-S}), so that eliminating the second component in the core becomes harder for higher temperature. We see that generally the instability of $p=2$ channel appears earlier than the instability of $p=3$ channel as $\nu$ increases. Whereas for large $\nu$ above the red lines of Figure~\ref{S=3}, the spectrum is identical to that of vortex in single component superfluids, since the vortex core is empty ($\Psi_2=0$), see Figure~\ref{O2-nu}. In the parameter range shown in Figure~\ref{S=3}, we find no instability for $p=4$ channel. This is because instability for $p=4$ channel only appears at high temperatures~\cite{RN393}. However, those high temperatures develop no instability for both $p=2$ and $p=3$ channels.

As is mentioned above, we are considering the immiscible binary superfluids (\emph{i.e.} $\nu>0$). Because of the repulsion between the two condensates, the second condensate $\mathcal{O}_2$ lies in the core of the vortex of the first one (see the left panel of Figure~\ref{O-S}) and acts as a stabilizer. Therefore, one expects the symbiotic vortex-bright soliton structure should be stable as in weakly coupling cases~\cite{RN349,RN350,RN365}, at least when $\mathcal{O}_2$ in the core is large enough. This feature does happen in our holographic theory. As clear from Figure~\ref{O2-nu} and Figure~\ref{S=3}, at each temperature, the quantized composite vortex is stable until the inter-component coupling  $\nu$ is sufficiently strong. 

\subsection{Full non-linear numerical simulations}
After revealing the occurrence of dynamical instability through linear quasi-normal mode analysis, we turn to the real time process of our quantized vortex by implementing full non-linear numerical simulations of the (3+1) dimensional bulk dynamics. The simulation starts with a static vortex-bright soliton, perturbed by some random noise.
\begin{equation}
    \Psi_i=\Psi_{0i}(1+\sum_{k=1}^N[\alpha(k)e^{-ik\theta}+\beta(k)e^{ik\theta}])\,,
\end{equation}
where $\alpha(k)$ and $\beta(k)$ are random small numbers, and $N$ is a truncation integer that is set to be 20. The details of time evolution scheme can be found in Appendix~\ref{A}. 
\begin{figure}[htpb]
        \centering
            \includegraphics[width=0.95\linewidth]{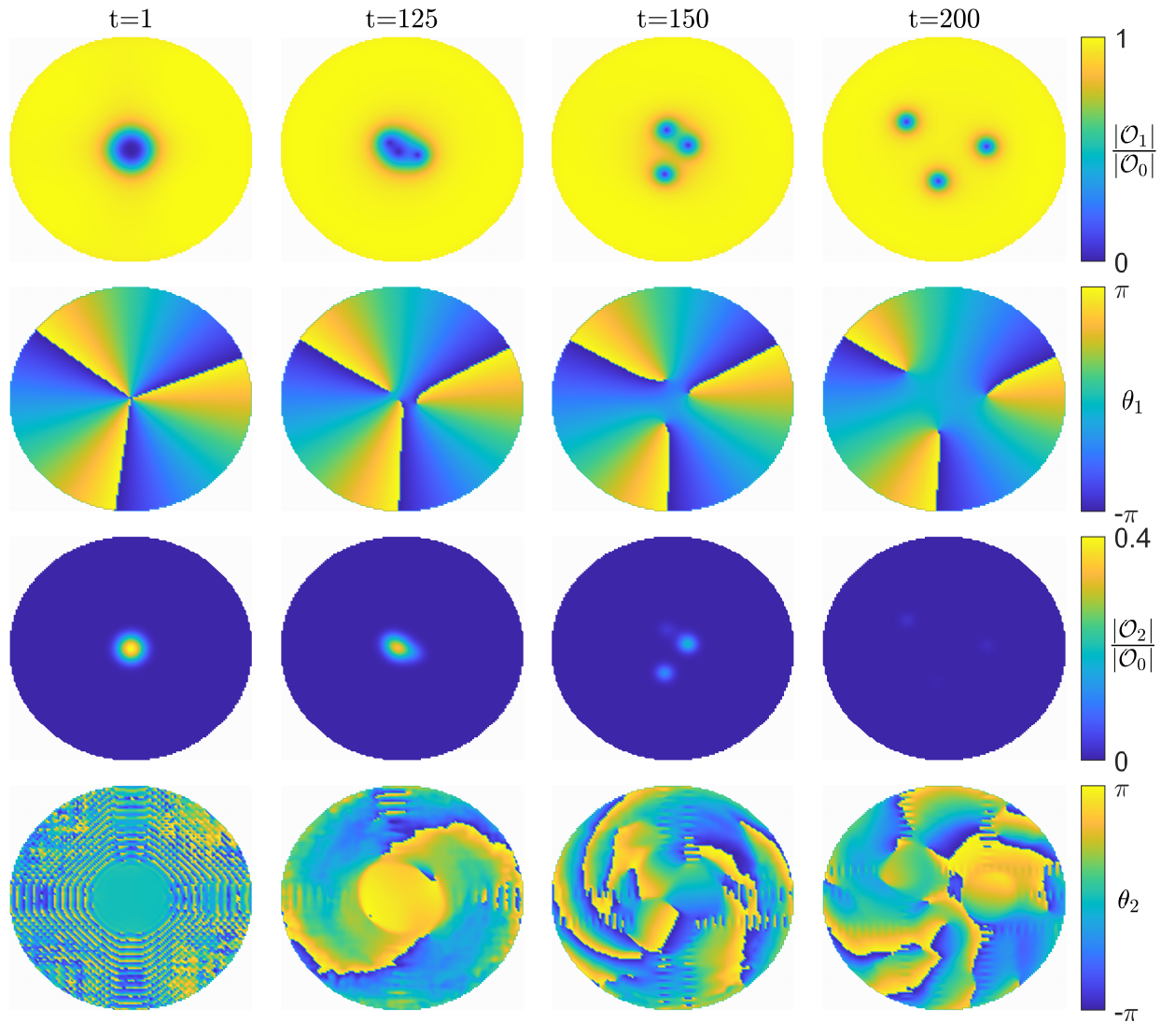}
            \caption{ Time evolution of the amplitude and phase profiles for symbiotic vortex-bright soliton with winding number $S=3$ at $T/T_c=0.677$ and $\nu=71$. From Top to bottom, plotted are $|\mathcal{O}_1|$, $\theta_1$, $|\mathcal{O}_2|$ and $\theta_2$. $|\mathcal{O}_0|$ is the value of the first order parameter far from the vortex core. Plotted region is $[-8,8]\times[-8,8]$. Noise in plots of $\theta_2$ is due to the fact that the value of $\mathcal{O}_2$ is too small to give reliable information of phase numerically. }
    \label{tevol}
\end{figure}

Figure~\ref{tevol} depicts $S=3$ composite vortex splitting at $T/T_c=0.677$ and $\nu=71$. From Top to bottom, plotted quantities are $|\mathcal{O}_1|$, $\theta_1$, $|\mathcal{O}_2|$ and $\theta_2$. Four snapshots are shown for each quantity from left to right. The locations of a vortex can be directly seen from the singularities or branch points of the phase configuration. The phase difference from blue to yellow is $2\pi$, which corresponds to a single vortex. There can be a large winding number vortex $|S|>1$ for which circling around such vortex core, the change of phase is $2S\pi$. At this particular temperature and coupling strength, $p=2$ channel should dominate the instability, based on our linear instability analysis shown in Figure~\ref{O2-nu}. From the plot at $t=125$, one can see the splitting instability indeed presents $p=2$ pattern. As time goes by, the $S=3$ vortex eventually splits into three $S=1$ vortices (see plot at $t=150$). However, in the present case, $S=1$ vortex only admits empty core. Therefore, the residual second component in each core is gradually eliminated while moving together with the vortices in the first component. As can be seen from the plots at $t=150$ and $t=200$, the maximum value of $|\mathcal{O}_2|$ decreases rapidly. 

Surprisingly, we find no instability for larger winding numbers. This again can be explained by the enlargement of the vortex radius. As visible from the right panel of Figure~\ref{O-S}, the size of the composite vortex grows fast as $S$ increases. When $|S|>3$, the radius of the vortex becomes so large that the second condensate can never be eliminated no matter how large $\nu$ is. To help the readers get some feeling, max$|\mathcal{O}_2|\simeq 5.9$ persists even to $\nu=10^5$ for $S=4$ at $T/T_c=0.677$, whereas max$|\mathcal{O}_1|\simeq 6.4$. For comparison, instability for $S=3$ composite vortex begins to occur when max$|\mathcal{O}_2|\lesssim 5$ at this temperature. From above discussion, it appears that as long as the second condensate in the core is strong enough, the symbiotic vortex-bright soliton structure is stable, which is the case for large winding number, low temperature and small coupling strength. 
Nevertheless, when the winding number is so large that the radius of the vortex is much larger than the interface width between the two superfluid condensates, the interface dynamics can trigger some instability around the interface. As we shall show in the next section, there would be instability due to the relative velocity between the two components, even though the second component in the core is never eliminated.

\section{Giant symbiotic vortex-bright soliton \& KHI}
\label{sec5}
In classical and quantum fluid, when there is relative velocity difference across the interface between two fluids, one kind of interface instability called KHI occurs. Such instability typically leads to roll-up patterns in the nonlinear stage. KHI of straight interface in holographic two-component superfluids has been studied in~\cite{2comKHI,An:2024dkn}.
As is mentioned above, the radius of symbiotic vortex-bright soliton grows fast as its winding number increases (see the right panle of Figure~\ref{O-S}). When the radius of the vortex is much larger than the interface width between the two superfluid condensates, such giant vortex can be regard locally as a two-fluid system sharing a straight interface with a relative velocity $v=S/R$, where $S$ is the winding number and $R$ is the vortex radius. Therefore, the develop of KHI is anticipated. 

KHI is expected to occur when the relative velocity across the interface exceeds critical value, which is proportional to external stabilizing force~\cite{QKH3}. For giant vortex-bright soliton without external potential or force, the centrifugal force is stabilizing the interface. However, if the vortex is large enough, this force is negligible and the interface can be considered as planar. For planar interface without external force, the mean-field approach predicts that such KHI triggers no matter how small the relative velocity is~\cite{RN312}. In contrast to the mean-field analysis, the holographic study suggests that such instability only occurs when the relative velocity is beyond a threshold value~\cite{2comKHI}. Therefore, giant vortex-bright soliton eventually becomes unstable as winding number increases in mean-field approach, while it remains stable for large winding numbers in our holographic model, as we will see.  
It appears that the system described by holography has a stabilization mechanism, from which one anticipates the occurrence of instability for giant vortex-bright soliton only above a critical relative velocity. Moreover, the critical relative velocity is independent of the winding number. 

\begin{figure*}[htpb]
        \centering
         \vspace{-0.35cm}
            \includegraphics[width=0.46\linewidth]{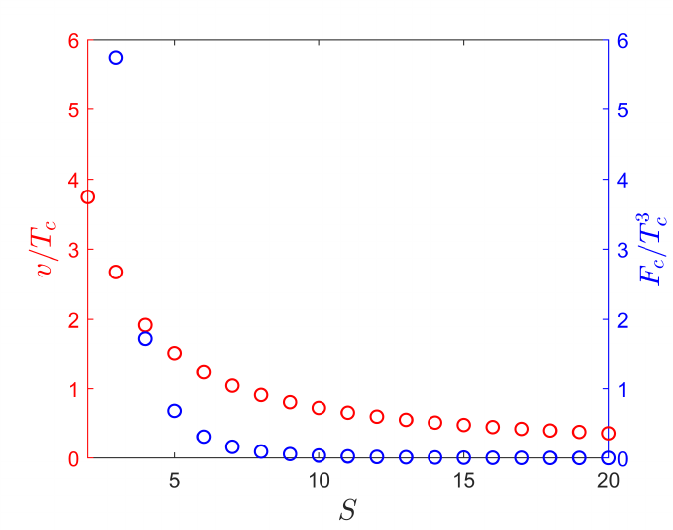}\;
\includegraphics[width=0.49\linewidth]{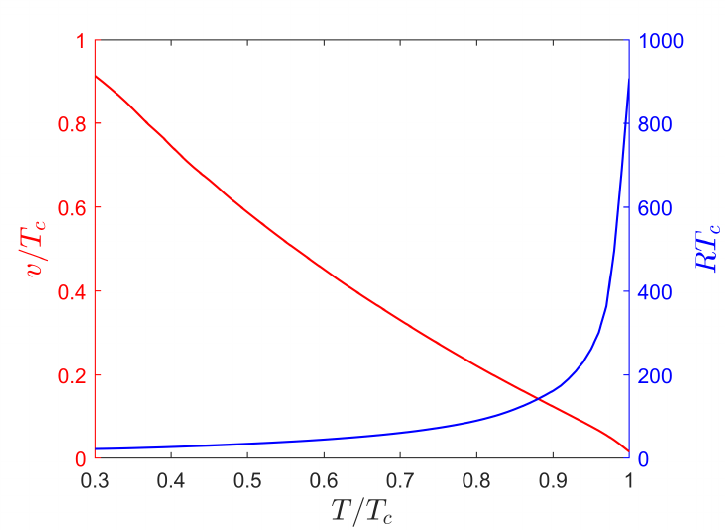}
            \caption{ \textbf{Left panel:} Effective relative velocity $v$ 
 and centrifugal force $F_c$ versus the winding number $S$ at $T/T_c=0.677$. \textbf{Right panel:} Effective relative velocity $v$ 
 and vortex radius $R$ with respect to the temperature $T$ for $S=20$.  We have taken $\nu=1$. }
    \label{R-v-S}
\end{figure*}

In the left panel of Figure~\ref{R-v-S}, we show the effective relative velocity $v=S/R$ versus the winding number $S$ at $T/T_c=0.677$, where the radius is estimated by $|\mathcal{O}_1(R)|=0.5\,\mathrm{max}|\mathcal{O}_1|$ for the stationary $S$-quantum vortex state.\,\footnote{The vortex core size $R$ is usually estimated by fitting the vortex order parameter as~\cite{superfluidbook} $\mathcal{O}_1(r)= \mathcal{O}_1(R)\tanh[r/(\sqrt{2}R)]$. Unfortunately, this formula fails for the giant vortex we consider. Therefore, we define the radius as $|\mathcal{O}_1(R)|=0.5\,\mathrm{max}|\mathcal{O}_1|$.} One can see that $v$ decreases quickly as $S$ is increased. Moreover, we show the centrifugal force $F_c\equiv v^2/R=S^2/R^3$ versus $S$. This force decreases fast as winding number increases. Therefore, for large winding number, the centrifugal force can be ignored. In the right panel of Figure~\ref{R-v-S} we give the temperature dependence of $v$ and $R$ for $S=20$, from which one can see that $v$ decreases ($R$ increases) monotonically with increasing $T$. For comparison, KHI begins to appear for straight interface when $v/T_c\gtrsim 1$ at $T/T_c=0.677$ and $\nu=1$~\cite{2comKHI}. Unfortunately, $v/T_c$ is not significantly larger than 1 even for moderate $S$ for the present model, and indeed, we observe no instability for $S\geqslant4$. Note that in the rotating case we are considering here, centrifugal force would push the critical velocity to be even larger. Therefore, in sharp contrast to the mean-field description of superfluidity~\cite{RN312}, we demonstrate that the formation of stable giant symbiotic vortex-bright solitons is a robust prediction of holography. 

For $S<4$, the vortex size is comparable to interface width and definition for radius of vortex used above is in fact not appropriate any more. One would expect the vortex-bright solition instability is no longer dominated by KHI for such small vortices. And indeed, we can see the centrifugal force becomes very large for small winding numbers. KHI should be suppressed by such large stabilizing force. Nevertheless, vortex-bright soliton instability still occurs, which should be attributed to vortex splitting instability. For single component vortices, splitting instability is presented generally~\cite{RN400,RN399,RN401,RN356,RN393}. Such instability was argued to be caused by negative energy modes of vortices~\cite{RN400}, and might have potential connection with black hole superradiance instability~\cite{RN408,RN429}. For immiscible binary superfluids, the second component in the vortex core of the first component seems to cause the transition from splitting instability to KHI as the winding number increases.

\subsection{KHI for non-identical superfluid systems}
Thus far, we have focused on the binary superfluids where the two superfluid components are identical. We find no instability for giant symbiotic vortex-bright solitons. Nevertheless, it does not mean there is no instability for other giant composite vertices. 

According to the above discussion, to observe the KHI in rotating case, we should increase the relative velocity, or equivalently, decrease the vortex radius. One way to achieve this aim is to adjust the charge of the two components, which can give rich competing and coexisting behaviors in the uniform case~\cite{RN404,RN403}. In practice, we keep $e_1=1$ and change $e_2$. According to the argument of~\cite{RN404}, in the probe limit with $m_1=m_2$, if $e_2<e_1$, the first component would always prevail over the second one in the homogeneous case. Vortex-bright solitons can be constructed following the procedure in Section~\ref{sec3}. Then we can extract the vortex radius $R$ and the effective relative velocity $v$.

\begin{figure*}[htpb]
        \centering
         \vspace{-0.35cm}            \includegraphics[width=0.48\linewidth]{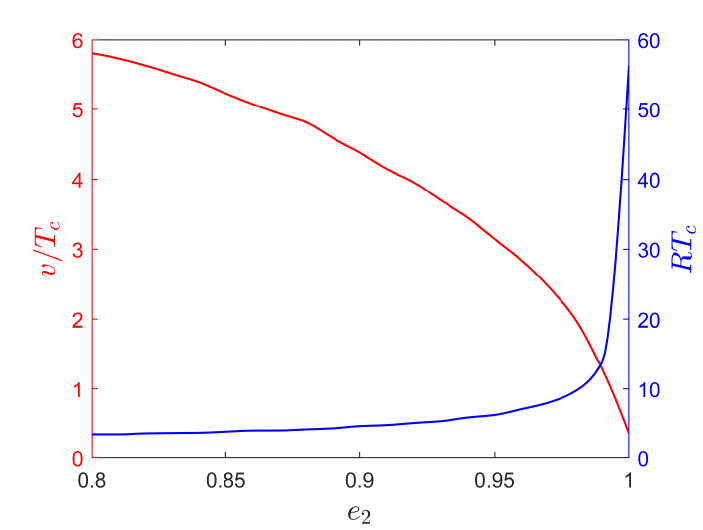}\;
            \includegraphics[width=0.49\linewidth]{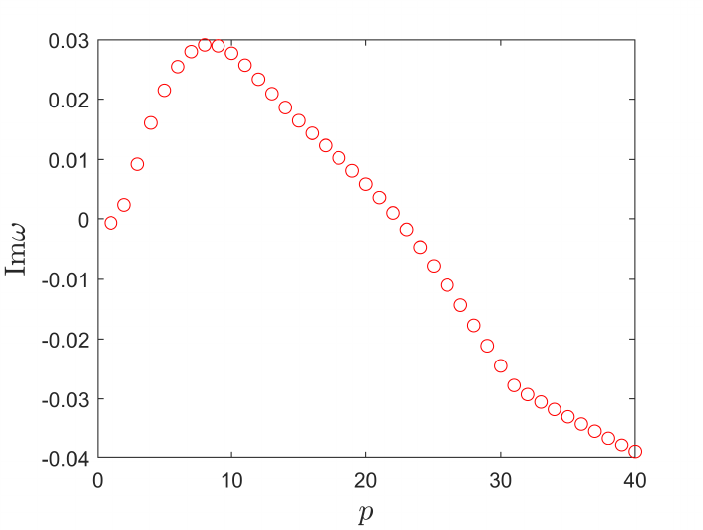}
            \caption{Illustration of physical variables extracted from symbiotic vortex-bright solitons for non-identical binary superfluids.   \textbf{Left panel:} The vortex radius $R$ and the corresponding effective relative velocity $v$ versus $e_2$. As $e_2$ decreases, $R$ decreases and $v$ increases rapidly. \textbf{Right panel:} Imaginary part of quasi-normal modes for $e_2=0.92$. The excitation mode has $p$-fold symmetry. The dominating mode is $p=8$. We have chosen $S=20$, $T/T_c=0.677$ and $\nu=1$ in both plots.}
    \label{R-v-e}
\end{figure*}

In the symbiotic vortex-bright soliton we are considering here, the second component $\mathcal{O}_2$ could still exist in the vortex core of the first one since $\mathcal{O}_1$ is always zero there. Nevertheless, by lowering $e_2$, the compelling force of the second component is effectively lowered, and therefore the vortex radius is decreased. Left panel of Figure~\ref{R-v-e} shows $R$ and $v$ versus $e_2$ for the winding number $S=20$ with $T/T_c=0.677$ and $\nu=1$. It's clear that by lowering $e_2$, $R$ decreases drastically and therefore $v$ is increased to be far above the critical velocity. Numerically we do find a dynamical instability as soon as $e_2<0.98$. 

\begin{figure}[htpb]
        \centering
            \includegraphics[width=0.95\linewidth]{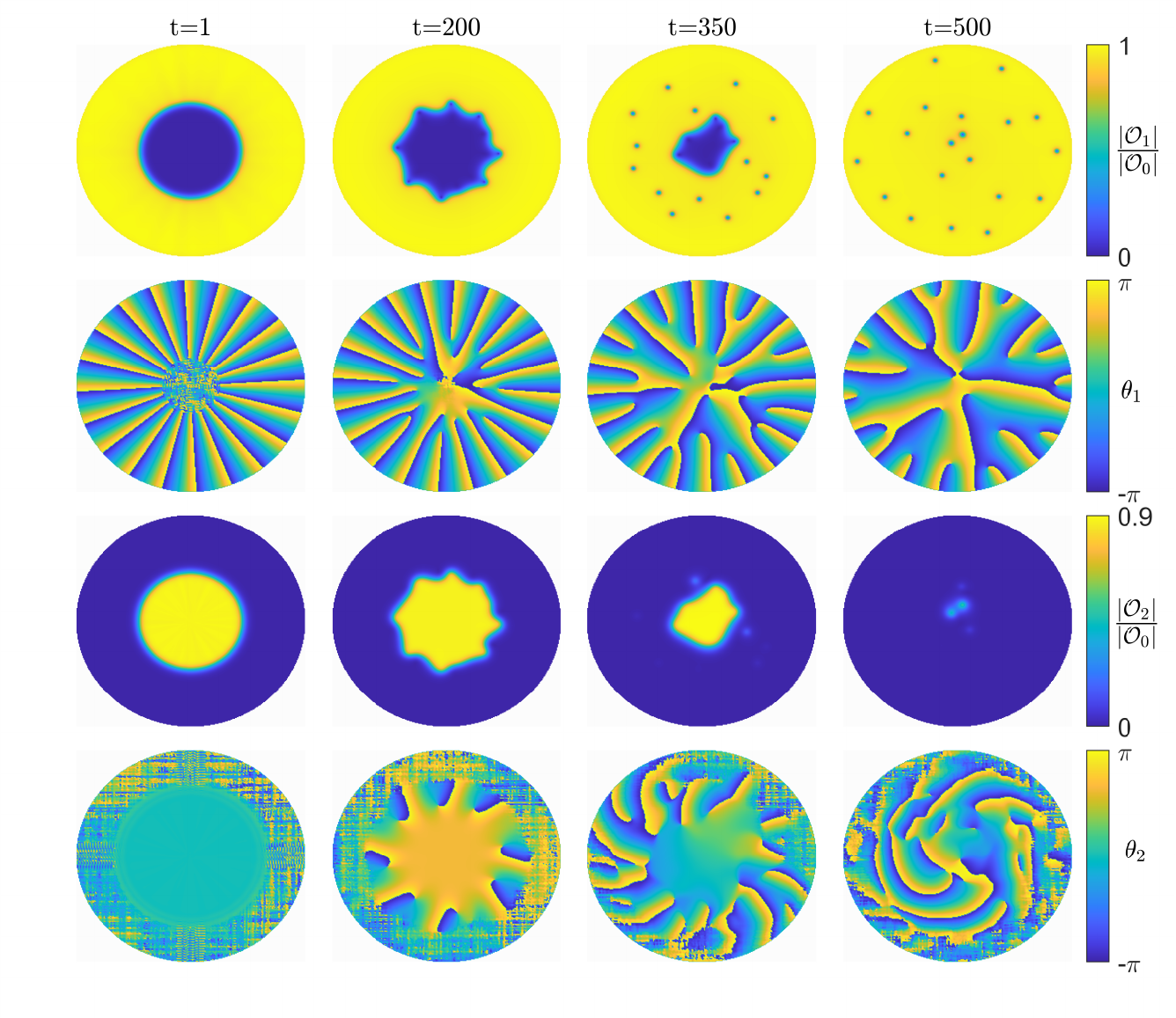}
            \caption{Time evolution of symbiotic vortex-bright soliton for non-identical binary superfluids with $e_2=0.92$, $S=20$, $T/T_c=0.677$ and $\nu=1$. Quantities plotted are the same as those in Figure~\ref{tevol}. Plotted region is $[-28,28]\times[-28,28]$. Early stage of the instability shows eight-fold pattern, in agreement with the linear analysis result that $p=8$ mode dominates. As the composite giant vortex splits into $S=1$ vortices, the second component is gradually eliminated.  Noise in plots of $\theta_2$ is due to the fact that value of $\mathcal{O}_2$ is too small to give physical information of phase.}
    \label{t_S=20}
\end{figure}

As an example, the spectrum solved from linear analysis for $e_2=0.92$ is presented in right panel of Figure~\ref{R-v-e}. One can see that there is a bell curve from which dynamical instability develops for $p=2, 3, \cdots$, 22. The imaginary part is a maximum for $p=8$ that correspond to 
the fastest growing mode. The full time evolution with the same parameters can be found in Figure~\ref{t_S=20}. The evolution scheme is the same as that used for $S=3$ case of Figure~\ref{tevol}, but with $N$ set to be 40. It is manifest that the core size is much larger than the width of the interface ($t=1$ in Figure~\ref{t_S=20}). Therefore, one can locally approximate this composite giant vortex as a two-fluid system sharing a straight interface with a relative velocity $v$. In the beginning ($t=1$), the second component $\mathcal{O}_2$ stays still at the vortex core, while the first component $\mathcal{O}_1$ is circulating with the velocity $v(r)=S/r$ around it.

As time goes by, the interface instability similar to KHI begins to develop. At $t=200$, a wavy pattern develops at the interface due to the KHI. The eight-fold pattern is manifest, which is in agreement with the result from linear analysis which shows $p=8$ channel dominates the instability. At the same time, the component $\mathcal{O}_2$ gradually begins to rotate with the first one. In particular, one can see that the relevant phase singularities of the second component fall within regions where the superfluid condensate $\mathcal{O}_2$ is almost zero. This induces the production of the so-call ``ghost vortices"~\cite{RN413}, which is unambiguously shown in subplot of phase $\theta_2$ at $t=200$. As the system evolves further, the wave at the interface then grows and quantized vortices of $\mathcal{O}_1$ with $S=1$ are released from the interface ($t=350$). This interface dynamics is similar to that in the flat interface~\cite{2comKHI}. Meanwhile, the condensate $\mathcal{O}_2$ in the core is eliminated by these $S=1$ vortices and the core size shrinks. This mechanism continues until the core totally disappears and the giant composite vortex with $S=20$ eventually splits into 20 $S=1$ vortices. From this picture, for a giant symbiotic vortex-bright soliton, the vortex splitting instability and KHI seems to be closely connected with each other.

\section{Conclusion and discussion}
\label{sec6}
In this paper, we have studied the (in)stability splitting dynamics of symbiotic vortex-bright soliton structure in holographic binary superfluids. It is a kind of composite vortex with the vortex core filled with another immiscible species. The second component in the vortex core generally acts as a stabilizer so as to suppress or even prevent vortex splitting instability. We have mainly focused on the case with identical superfluids. We find that for the winding number $S\leqslant 3$, the second component can be eliminated from the vortex core by increasing the coupling strength $\nu$ between the two components. When the core is ``empty enough", the splitting instability appears again. However, for $S\geqslant 4$, the second component $\mathcal{O}_2$ in the core can never be eliminated no matter how large $\nu$ is. This is attributed to the fact that the vortex radius grows fast as the winding number increases. Larger vortex core gives larger region for $\mathcal{O}_2$ to exist, and makes it harder to clean the core. And therefore we find no instability for the composite vortex for $S\geqslant 4$.

When the winding number is large, \emph{e.g.} $S\sim 10$, the radius of $S$ quantum vortex would be much larger than the interface width between the two condensates. Such giant vortices can be regarded as two-fluid systems with relative velocity $v=S/R$, where $R$ is the vortex radius, for which the KHI is expected for large enough $v$, even though the second component is as strong as the first component circulating it. Unfortunately, numerical results for $e_1=e_2$ cases show that $v$ is always too small to produce the KHI. To achieve a large relative velocity, we have considered a smaller value of $e_2$, which effectively lowers the size of the second component and decreases the vortex radius drastically, see Figure~\ref{R-v-e}.
By this means, we have managed observing the KHI in such composite giant vortices. As shown in Figure~\ref{t_S=20}, the final state of such instability is the total split of the giant vortex and many singly quantized vortices left behind. This implies a close connection of vortex splitting instability and KHI in two-component superfluids. 
\begin{figure}[htpb]
        \centering
            \includegraphics[width=0.95\linewidth]{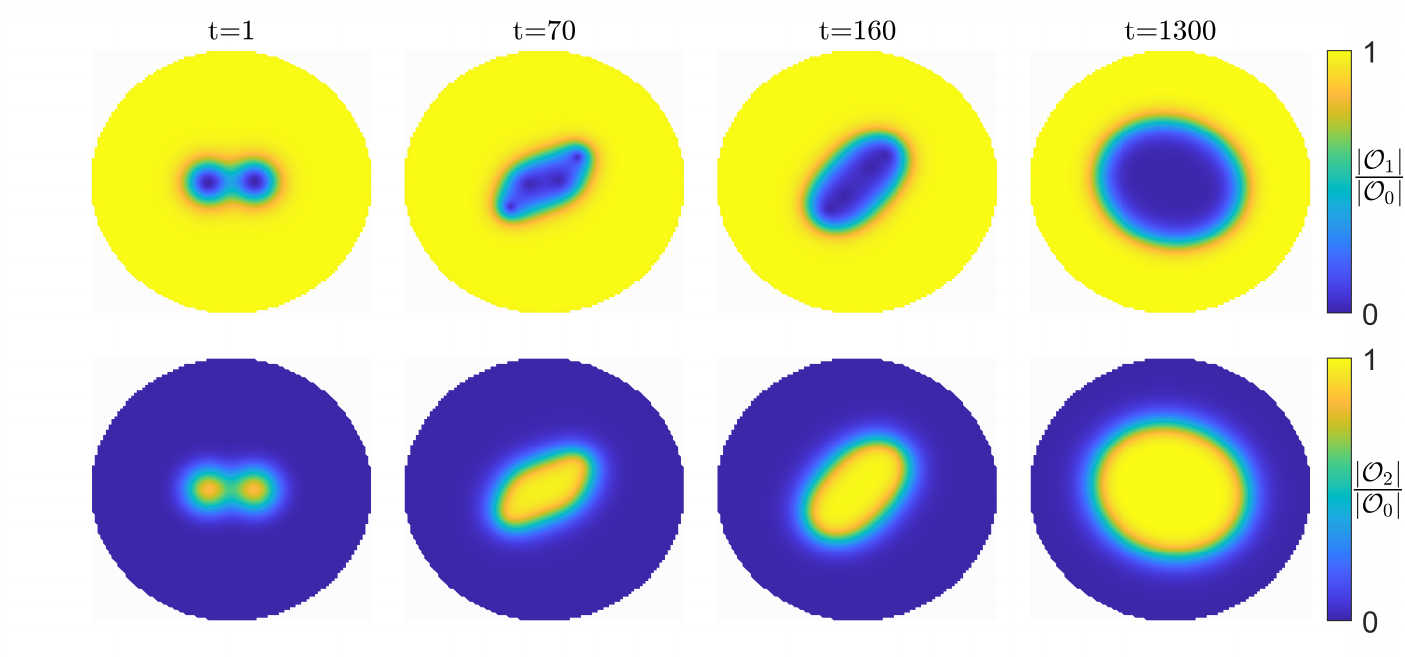}
            \caption{Two $S=2$ vortices merge into a single $S=4$ vortex at $T/T_c=0.677$, $\nu=1$ and $e_1=e_2$. The final state is a symbiotic vortex-bright soliton with $S=4$. Plotted region is $[-25,25]\times[-25,25]$. }
    \label{merge}
\end{figure}

There are several possible directions to extend our work. As shown from the left panel of Figure~\ref{R-v-S}, the effective relative velocity decreases as the winding number increases for $e_1=e_2$. This implies that vortex with larger winding number is more stable, therefore opens the possibility for vortices with smaller winding number to merge into vortices with larger winding number. In Figure~\ref{merge} we give an example for two $S=2$ vortices merging into a symbiotic vortex-bright soliton with $S=4$. This merging process is in sharp contrast to the case in single-component superfluids for which the vortices rotating in the same direction always repel each other. Exploring consequence of such spontaneous merging mechanism in strongly interacting binary superfluid systems, such as neutron stars~\cite{Haskell2018}, is of great importance. The wave number of the fastest growing modes yields a non-monotonic dependence of the relative velocity for straight interface~\cite{2comKHI}, it would be interesting to investigate whether similar non-monotonic feature exists in this rotating case. We have limited ourselves to the case for which the second superfluid component $\mathcal{O}_2$ has zero winding number. Turning on a finite winding number of $\mathcal{O}_2$ will enrich the structure of vortex-bright solitons and will introduce more interesting splitting and merging patterns and dynamical transitions. In particular, interface instability for two superfluid components moving with identical velocity was observed recently~\cite{An:2024dkn}, one may desire if this instability can be found in the co-rotating composite vortices. Those studies can be relevant not only for superfluids in the astrophysical context, but also for superfluids in the laboratory such as cold atomic gases. It is interesting to understand the physical mechanism behind the instability of symbiotic vortex-bright solitons in this model.

\appendix
\section{Equations of motion and time evolution scheme}
\label{A}
The Lagrangian~\eqref{action} yields the equations of motion for $\Psi_i$ and $A_\mu$:
 \begin{equation}
    \mathcal{D}_\mu \mathcal{D}^\mu\Psi_i-m_i^2\Psi_i-\frac{\nu}{2}|\Psi_j|^2\Psi_i=0\,, \quad(i,j=1,2\quad i\ne j),
 \end{equation}
 \begin{equation}
     \nabla_\mu F^{\mu\nu}=-2\mathrm{Im}(\sum_i\Psi^*_i\mathcal{D}^\nu\Psi_i)\,,
 \end{equation}
where $\mathrm{Im}$ represents the imaginary part. While it is convenient to consider vortex-bright soliton structures using polar coordinates, we choose rectangular coordinates for the time evolution to achieve better numerical accuracy.
 \begin{equation}
        ds^2=\frac{L^2}{z^2}(-f(r)dt^2-2dtdz+dx^2+dy^2)\,.
    \end{equation}
All functions in rectangular coordinates can be easily transformed to polar coordinates via
    \begin{equation}
        \begin{aligned}
            x=r\mathrm{cos}\theta,& \quad y=r\mathrm{sin}\theta, \\
            A_x=A_r\mathrm{cos}\theta-\frac{A_\theta}{r}\mathrm{sin}\theta,&\quad
            A_y=A_r\mathrm{sin}\theta+\frac{A_\theta}{r}\mathrm{cos}\theta\,.
        \end{aligned}
    \end{equation}

The explicit form of bulk equations of motion in rectangular coordinates reads
    \begin{equation}
        \begin{aligned}
            \label{phi}
            2\partial_t\partial_z\Phi_i-[2i A_t\partial_z\Phi_i+i \partial_zA_t\Phi_i+\partial_z(f\partial_z\Phi_i)-z\Phi_i
            +\partial_x^2\Phi_i+\partial_y^2\Phi_i
            -i (\partial_xA_x+\partial_yA_y)\Phi_i&\\
            -(A_x^2+A_y^2)\Phi_i-2i (A_x\partial_x\Phi_i+A_y\partial_y\Phi_i)
            -\frac{\nu}{2}|\Phi_j|^2\Phi_i]=0\,, \qquad(i,j=1,2,\quad i\ne j)&
        \end{aligned}
    \end{equation}
    \begin{equation}
        \label{At}
        \begin{aligned}
            \partial_t\partial_zA_t-[\partial_x^2A_t+\partial_y^2A_t+f\partial_z(\partial_xA_x+\partial_yA_y)-\partial_t(\partial_xA_x+\partial_yA_y)
            -2A_t\sum_i|\Phi_i|^2&\\
            -2f\mathrm{Im}(\sum_i\Phi_i^*\partial_z\Phi_i)+2\mathrm{Im}(\sum_i\Phi_i^*\partial_t\Phi_i)]=0\,,&
        \end{aligned}
    \end{equation}
    \begin{equation}
        \label{Ax}
        \begin{aligned}
            2\partial_t\partial_zA_x-[\partial_z(\partial_xA_t+f\partial_zA_x)+\partial_y(\partial_yA_x-\partial_xA_y)-2A_x\sum_i|\Phi_i|^2
            +2\mathrm{Im}(\sum_i\Phi_i^*\partial_x\Phi_i)]=0\,,
        \end{aligned}
    \end{equation}
    \begin{equation}
        \label{Ay}
        \begin{aligned}           
        2\partial_t\partial_zA_y-[\partial_z(\partial_yA_t+f\partial_zA_y)+\partial_x(\partial_xA_y-\partial_yA_x)-2A_y\sum_i|\Phi_i|^2
            +2\mathrm{Im}(\sum_i\Phi_i^*\partial_y\Phi_i)]=0\,,
        \end{aligned}
    \end{equation}
        \begin{equation}
        \label{constraint}
        \begin{aligned}     
        \partial_z(\partial_xA_x+\partial_yA_y-\partial_zA_t)-2\mathrm{Im}(\sum_i\Phi_i^*\partial_z\Phi_i)=0\,,
        \end{aligned}
    \end{equation}
where $\Phi_i=\Psi_i/z$. Notice the last equation is a constraint with no time derivative. These equations are not independent. More precisely, they obey the following constraint equation:
\begin{equation}
    \label{relation}
        -\partial_t\mathrm{Eq.}(\ref{constraint})-\partial_z\mathrm{Eq.}(\ref{At})+\partial_x\mathrm{Eq.}(\ref{Ax})+\partial_y\mathrm{Eq.}(\ref{Ay})=2\mathrm{Im}(\sum_i\mathrm{Eq.}(\ref{phi})\times\Phi_{0i}^*)\,.
    \end{equation}

The initial condition is chosen to be a stationary vortex bright soliton perturbed by some random noise 
    \begin{equation}
    \Psi_i=\Psi_{0i}\left(1+\sum_{k=1}^N[\alpha(k)e^{-ik\theta}+\beta(k)e^{ik\theta}]\right),
    \end{equation}
where $\alpha(k)$ and $\beta(k)$ are some random small numbers, and $N$ is set to be 20 for $S=3$ and 40 for $S=20$\,.  

For time evolution, we use the fourth order Runge-Kutta method and the following scheme. First, we use \eqref{phi}, \eqref{Ax} and \eqref{Ay} to evolve $\Phi_i$, $A_x$ and $A_y$ with boundary conditions $\Phi_i(z=0)=A_x(z=0)=A_y(z=0)=0$. Then we use \eqref{At} to evolve $\partial_zA_t$ on the boundary. Note that $-\partial_zA_t(z=0)$ is just the charge density or number density $\rho$ of the dual field theory. Finally, we use \eqref{constraint} to solve $A_t$ by evolved $\Phi_i$, $A_x$, $A_y$ and boundary conditions $\partial_zA_t(z=0)=-\rho$ and $A_t(z=0)=\mu$. Such scheme keeps the chemical potential $\mu$, unchanged, so we are in fact working in grand canonical ensemble. In $z$ direction, we use Chebyshev pseudo spectral method. In both $x$ and $y$ directions we use the Fourier pseudo spectral method and the periodic boundary condition. In practice, to apply periodic boundary condition for vortex, one can prepare a pancake shape so that field values are vanishing on boundaries. Following~\cite{RN356}, such configuration is achieved by setting
\begin{equation}
A_t|_{z=0}=\mu/2(1-\mathrm{tanh}[c(r^2-r_m^2)])\,,
\end{equation}
with $r_m$ and $c$ constants. Moreover, the value of $r_m$ should be sufficiently large compared to the vortex size such that the intrinsic vortex dynamics is not affected. And typically, $c$ is set to be $0.05$.

\section{Numerical details for quasi-normal modes of quantized vortices}
\label{B}
The linearized equations of perturbations~\eqref{pfold} are explicitly given as follows.  For convenience, below we omit the subscript 0 of background configurations.
\begin{eqnarray}\label{u}
&&(-2i(\omega+A_{t})\partial_{z}-i\partial_{z}A_{t}-\partial_{z}(f\partial_{z})-(\partial_{r}-i A_{r})^{2}-\frac{\partial_{r}-i A_{r}}{r}+\frac{(A_{\theta}-S_i-p)^{2}}{r^{2}}+z+\frac{\nu}{2}|\Phi_j|^2)u_i\nonumber\\
&&+\frac{\nu}{2}\Phi_j^*\Phi_iu_j+\frac{\nu}{2}\Phi_j\Phi_iv_j -(i \Phi_i\partial_{z}+2i\partial_{z}\Phi_i)a_t+(i\Phi_i\partial_{r}+2i\partial_{r}\Phi_i+2\Phi_i A_{r}+\frac{i\Phi_i}{r})a_r\nonumber\\
&&+\frac{\Phi_i}{r^{2}}(2A_{\theta}-2S_i-p)a_\theta=0,\quad (i,j=1,2,\quad i\ne j)\,,
\end{eqnarray}
\begin{eqnarray}\label{v}
&&(-2i(\omega-A_{t})\partial_{z}+i\partial_{z}A_{t}-\partial_{z}(f\partial_{z})-(\partial_{r}+i A_{r})^{2}-\frac{\partial_{r}+i A_{r}}{r}+\frac{(A_{\theta}-S_i+p)^{2}}{r^{2}}+z+\frac{\nu}{2}|\Phi_j|^2)v_i\nonumber\\
&&+\frac{\nu}{2}\Phi_j^*\Phi_i^*u_j+\frac{\nu}{2}\Phi_j\Phi_i^*v_j +(i \Phi_i^{*}\partial_{z}+2i \partial_{z}\Phi_i^{*})a_t-(i \Phi_i^{*}\partial_{r}+2i\partial_{r}\Phi_i^{*}-2\Phi_i^{*}A_{r}+\frac{i \Phi_i^{*}}{r})a_r\nonumber\\
&&+\frac{\Phi_i^{*}}{r^{2}}(2 A_{\theta}-2S_i+p)a_\theta=0, \quad (i,j=1,2,\quad i\ne j)\,,
\end{eqnarray}
\begin{eqnarray}\label{da1}
    \sum_i[(-i\Phi_i^{*}\partial_{z}+i\partial_{z}\Phi_i^{*})u_i+(i\Phi_i\partial_{z}-i\partial_{z}\Phi_i)v_i]+\partial_{z}^{2}a_t-(\frac{\partial_{z}}{r}+\partial_{z}\partial_{r})a_r-\frac{i p}{r^{2}}\partial_{z}a_\theta=0\,,
\end{eqnarray}
\begin{eqnarray}\label{a2}  &&\sum_i[(\omega\Phi_i^{*}+2A_{t}\Phi_i^{*})u_i+(-\omega\Phi_i+2A_{t}\Phi_i)v_i]+(\frac{p^{2}}{r^{2}}+2\sum_i|\Phi_i|^2-i\omega\partial_{z}-f\partial_{z}^{2}-\frac{\partial_{r}}{r}-\partial_{r}^{2})a_t\nonumber\\
  &&-(\frac{i\omega}{r}+i\omega\partial_{r})a_r+\frac{ p\omega}{r^{2}}a_\theta=0\,,  
\end{eqnarray}
\begin{eqnarray}\label{a3}
  &&\sum_i[(i\Phi_i^{*}\partial_{r}-i\partial_{r}\Phi_i^{*}+2\Phi_i^{*}A_{r})u_i+(-i\Phi_i\partial_{r}+i\partial_{r}\Phi_i+2\Phi_i A_{r})v_i]-\partial_{z}\partial_{r}a_t\nonumber\\
  &&+(\frac{p^{2}}{r^{2}}+2\sum_i|\Phi_i|^2-2i\omega\partial_{z}-f\partial_{z}^{2}-f'\partial_{z})a_r+\frac{ i p}{r^{2}}\partial_{r}a_\theta=0\,,   
\end{eqnarray}
\begin{eqnarray}\label{a4}
  &&\sum_i[(2A_{\theta}-2S_i-p)\Phi_i^{*}u_i+(2A_{\theta}-2S_i+p)\Phi_i v_i]-i p\partial_{z}a_t
  +i p(\partial_{r}-\frac{1}{r})a_r\nonumber\\
  &&+(-2i\omega\partial_{z}-\partial_{z}(f\partial_{z})-\partial_{r}^{2}+\frac{\partial_{r}}{r}+2\sum_i|\Phi_i|^2)a_\theta=0\,.
\end{eqnarray}
We use the gauge dependent formalism to calculate the quasi-normal modes by gauge fixing~\cite{Du:2015zcb}. In particular, the fourth equation~\eqref{a2} as a constraint equation will be used only at the AdS boundary $z=0$. To obtain the quasi-normal modes, we turn off all sources for perturbations at the AdS boundary.
\begin{eqnarray}
    u_i|_{z=0}=0,\,\,v_i|_{z=0}=0,\,\,a_t|_{z=0}=0,\,\,a_r|_{z=0}=0,\,\,a_\theta|_{z=0}=0\,.
\end{eqnarray}
At the event horizon $z=1$, the regular boundary conditions are imposed as usual. At the vortex core $r = 0$, the boundary conditions are determined by the asymptotic behavior
of the perturbation equations, for which we have
\begin{equation}
    u_i|_{r=0}=0,\quad v_i|_{r=0}=0,\quad a_t|_{r=0}=0,\quad a_\theta|_{r=0}=0,\quad (pa_r+i\partial_ra_\theta))|_{r=0}=0\,.
\end{equation}
At the cutoff $ r = R_0$, we impose the following boundary conditions:
   \begin{equation}
       \partial_ru_i|_{r=R_0}=0,\quad \partial_rv_i|_{r=R_0}=0,\quad \partial_ra_t|_{r=R_0}=0,\quad a_r|_{r=R_0}=0,\quad \partial_ra_\theta|_{r=R_0}=0\,.
   \end{equation}

\begin{figure}[htpb]
        \centering         \includegraphics[width=0.7\linewidth]{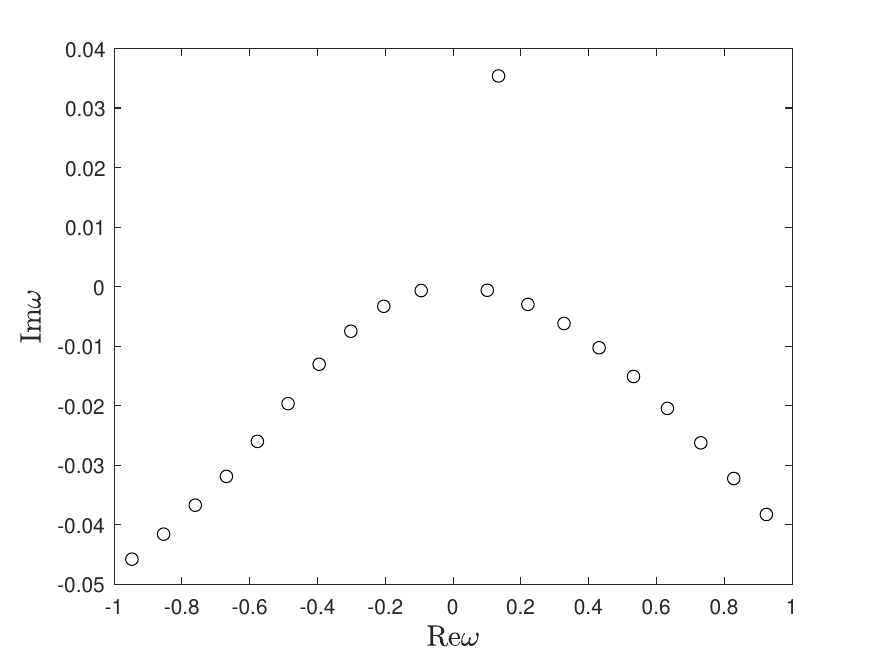}
            \caption{The low lying quasi-normal mode spectrum for $p=2$ of the symbiotic vortex-bright soliton with winding number $S=3$. Other parameters are $T/T_c=0.677$ ($\mu=6$), $\nu=71$. One single unstable mode with positive  imaginary part can be identified.}
    \label{example}
\end{figure}

We then arrive at a generalized eigenvalue problem. We can obtain the quasi-normal modes $\omega$ for each $p$ by numerically solving such generalized eigenvalue problem. The background configuration becomes linearly unstable whenever $\mathrm{Im}(\omega)>0$. The spectrum of quasi-normal mode for $p=2$ of the symbiotic vortex-bright soliton with winding number $S=3$ is depicted in Figure~\ref{example}. The fastest growing mode is given by a quasi-normal frequency with the largest positive imaginary part. There is only one frequency with positive imaginary part in Figure~\ref{example}, indicating the instability of the corresponding vortex-bright soliton for $p=2$ channel.

Notice that the complex conjugate of the above linearized equations can be obtained by the following transformation:
 \begin{equation}
     p\rightarrow -p,\quad \omega\rightarrow -\omega^*, \quad u_i\leftrightarrow v^*_i ,\quad a_\mu \rightarrow a_\mu^*\,.
 \end{equation}
Therefore, whenever $\omega$ is an eigenvalue for given $p$, $-\omega^*$ is an eigenvalue for $-p$, and they share the same imaginary part. Without loss of generality, we only need to consider channels with positive $p$.

\acknowledgments
This work was partly supported by the National Natural Science Foundation of China Grants No.\,12075298, No.\,12122513, No.\,11991052 and No.\,12047503. We acknowledge the use of the High Performance Cluster at Institute of Theoretical Physics, Chinese Academy of Sciences.


\bibliographystyle{JHEP}
\bibliography{biblio.bib}

\end{document}